\documentclass[number,11pt,1p]{elsarticle}

\begin{document}

\begin{frontmatter}



\title{Thunderstorm Observations by Air-Shower Radio Antenna Arrays}

\cortext[cor]{Corresponding authors}

\author[1]{W.D.~Apel}
\author[2,14]{J.C.~Arteaga}
\author[4]{L.~B\"ahren}
\author[1]{K.~Bekk}
\author[5]{M.~Bertaina}
\author[6]{P.L.~Biermann}
\author[1,2]{J.~Bl\"umer}
\author[1]{H.~Bozdog}
\author[7]{I.M.~Brancus}
\author[8]{P.~Buchholz}
\author[4,19]{S.~Buitink}
\author[5,9]{E.~Cantoni}
\author[5]{A.~Chiavassa}
\author[1]{K.~Daumiller}
\author[2,15]{V.~de Souza}
\author[1]{P.~Doll} 
\author[2,17]{M.~Ender\corref{cor}}\ead{Moses.Ender@kit.edu}
\author[1]{R.~Engel}
\author[4,10]{H.~Falcke} 
\author[1]{M.~Finger} 
\author[11]{D.~Fuhrmann}
\author[3]{H.~Gemmeke}
\author[8]{C.~Grupen}
\author[1]{A.~Haungs}
\author[1]{D.~Heck}
\author[4]{J.R.~H\"orandel}
\author[6]{A.~Horneffer}
\author[2]{D.~Huber\corref{cor}}\ead{Daniel.Huber@kit.edu}
\author[1]{T.~Huege}
\author[1,18]{P.G.~Isar}
\author[11]{K.-H.~Kampert}
\author[2]{D.~Kang}
\author[3]{O.~Kr\"omer}
\author[4]{J.~Kuijpers}
\author[2]{K.~Link}
\author[12]{P.~{\L}uczak}
\author[2]{M.~Ludwig}
\author[1]{H.J.~Mathes}
\author[2]{M.~Melissas}
\author[9]{C.~Morello}
\author[1]{S.~Nehls}
\author[1]{J.~Oehlschl\"ager}
\author[2]{N.~Palmieri}
\author[1]{T.~Pierog}
\author[11]{J.~Rautenberg}
\author[1]{H.~Rebel}
\author[1]{M.~Roth}
\author[3]{C.~R\"uhle}
\author[7]{A.~Saftoiu}
\author[1]{H.~Schieler}
\author[3]{A.~Schmidt}
\author[1]{F.G.~Schr\"oder}
\author[13]{O.~Sima}
\author[7]{G.~Toma}
\author[9]{G.C.~Trinchero}
\author[1]{A.~Weindl}
\author[1]{J.~Wochele}
\author[1]{M.~Wommer}
\author[12]{J.~Zabierowski}
\author[6]{J.A.~Zensus}

\address[1]{Karlsruhe Institute of Technology (KIT), Institut f\"ur Kernphysik, Karlsruhe, Germany}
\address[2]{Karlsruhe Institute of Technology (KIT), Institut f\"ur Experimentelle Kernphysik, Karlsruhe, Germany}
\address[4]{Radboud University Nijmegen, Department of Astrophysics, The Netherlands}
\address[5]{Dipartimento di Fisica Generale dell' Universita Torino, Italy}
\address[6]{Max-Planck-Institut f\"ur Radioastronomie Bonn, Germany}
\address[7]{National Institute of Physics and Nuclear Engineering, Bucharest, Romania}
\address[8]{Universit\"at Siegen, Fachbereich Physik, Germany}
\address[9]{INAF Torino, Istituto di Fisica dello Spazio Interplanetario, Italy}
\address[10]{ASTRON, Dwingeloo, The Netherlands}
\address[11]{Universit\"at Wuppertal, Fachbereich Physik, Germany}
\address[3]{Karlsruhe Institute of Technology (KIT), Institut f\"ur Prozessdatenverarbeitung und Elektronik, Karlsruhe, Germany}
\address[12]{Soltan Institute for Nuclear Studies, Lodz, Poland}
\address[13]{University of Bucharest, Department of Physics, Bucharest, Romania}

\address[14]{\scriptsize{now at: Universidad Michoacana, Morelia, Mexico}}
\address[19]{\scriptsize{now at: Lawrence Berkeley National Laboratory, United States}}
\address[15]{\scriptsize{now at: Universidade de S$\tilde{a}$o Paulo, Instituto de F\^{\i}sica de S$\tilde{a}$o Carlos, Brasil}}
\address[17]{\scriptsize{{now at: KIT, Institut f\"ur Werkstoffe der Elektrotechnik, Karlsruhe, Germany}}}
\address[18]{\scriptsize{{now at: ISS, Bucharest, Romania}}}

\begin{abstract}
Relativistic, charged particles present in extensive air showers lead to a coherent emission of 
radio pulses which are measured to identify the shower initiating high-energy cosmic rays. 
Especially during thunderstorms, there are additional strong electric fields in the atmosphere, 
which can lead to further multiplication and acceleration of the charged particles
and thus have influence on the form and strength of the radio emission. 
For a reliable energy reconstruction of the primary cosmic ray by means of the measured radio signal  
it is very important to understand how electric fields affect the radio emission. 
In addition, lightning strikes are a prominent source of broadband radio emissions that are visible 
over very long distances. 
This, on the one hand, causes difficulties in the detection of the much lower signal of the air shower. 
On the other hand the recorded signals can be used to study features of the lightning development. 
The detection of cosmic rays via the radio emission and the influence of strong electric 
fields on this detection technique is investigated with the LOPES experiment in Karlsruhe, Germany. 
The important question if a lightning is initiated by the high electron density given at the
maximum of a high-energy cosmic-ray air shower is also investigated, but could not be answered by LOPES. 
But, these investigations exhibit the capabilities of EAS radio antenna arrays for lightning studies. 
We report about the studies of LOPES measured radio signals of air showers
taken during thunderstorms and give a short outlook to 
new measurements dedicated to search for correlations of lightning and cosmic rays. 
\end{abstract}

\begin{keyword}
Cosmic Rays \sep Radio Detection \sep Thunderstorms \sep Lightning 
\PACS 96.50.sd \sep 87.50.sj \sep 52.80.mg
\end{keyword}

\end{frontmatter}


\section{Introduction}
For studying high-energy cosmic rays above $\approx 10^{15}$\,eV ground based observatories covering large areas 
are needed because the flux of the cosmic rays gets too low at these energies for direct measurements by balloon 
or satellite bound experiments. 
By ground based detection arrays instead, the secondary particles of an extensive air shower (EAS) generated by 
the impinging primary cosmic ray are measured. 
Beside the standard method of EAS detection with an array of particle detectors, one possibility to observe 
these secondary particles is to measure the radio emission that is produced by the electrons and positrons due 
to their deflection in the Earth's magnetic field. 
This radio emission gives an integrated measurement all over the shower development in the atmosphere 
contrary to particle detectors that only record a footprint of the shower at observation level. 
In contrast to the fluorescence or Cherenkov detection technique, where also the longitudinal shower development 
is observed by means of the EAS's electromagnetic emission, radio signals can be measured at night- and 
daytime~\citep{haungs2003}.
The radio detection technique can complement the information gained by particle detectors as well as can be 
operated as a stand alone detection technique. 
The problem of this technique lies in the weak signal compared to the 
ambient noise. But, with highly sensitive radio antenna arrays most of the noise sources can be 
identified and the data corrected for~\citep{SchroederArena}, 
except one of the most disturbing noise source formed by the radio emission of lightning strikes. 

Lightnings have been explored for a very long time, but the mechanism that leads to the discharging process 
is still not fully understood~\citep{dwyer}. 
There are several theories that could explain the sudden electric breakdown, 
where a very intriguing theory 
predicts that lightning strikes might be initiated by cosmic rays. 
To validate this theory, cosmic rays and lightning strikes need to be measured at the same place and time. 
As lightning observations are most effective in radio or X-ray due to the given transparency of the 
thunderclouds in these wavelengths, it seems natural to use EAS radio antenna arrays for such 
measurements.

The digital radio antenna array LOPES~\citep{Falcke05,Hau09} was designed and built to perform the
proof-of-principle of the EAS radio detection technique.  
LOPES is co-located and combined with the air-shower detector
KASCADE-Grande~\citep{kascade,grande}. 
KASCADE-Grande is a multi-detector air-shower experiment located 
in Karlsruhe, Germany, and consists of different particle-detector devices to measure all kinds of 
secondary particles in a large primary energy range of $10^{14}-10^{18}$\,eV.  

In order to study the effect of large atmospheric electric fields on the EAS radio signal  
present during thunderstorms, detailed investigations are performed at LOPES. 
It is shown that LOPES is able to detect air showers and, in addition, lightning strikes. 
Hence, LOPES can be used as a lightning mapping array and the correlation between EAS and 
lightning strikes is studied by getting the position and 
time of the lightning strike from LOPES and the air shower data from KASCADE-Grande. 
So far no correlation is seen, but with the technology used at LOPES the quality of the information 
as a lightning mapping array is rather low. 
However, by deploying lightning mapping systems that are particularly designed for lightning 
detection, see~\citep{krehbiel}, e.g., a far better time and spatial resolution of the lightning 
can be achieved. 
In addition, an increase of the lightning statistics would be desirable as the area covered by 
LOPES and KASCADE-Grande is small compared to the size of thunderstorm clouds. 
A future project will combine all these by deploying special 
designed lightning detectors at a large area cosmic ray surface experiment. 
By covering a huge area and having high quality information of 
both, lightning strikes and cosmic rays, one will be able to decide whether there is a 
correlation between lightning strikes and cosmic rays or not.

In this article we summarize the activities within LOPES with respect to thunderstorm observations and
briefly discuss the prospects of future projects.

\section{EAS Observations by Radio Antenna Arrays}
The measurement of cosmic ray induced air showers can be performed with different detection techniques. 
One of these techniques is the analysis of the radio pulses emitted by the electromagnetic component of an 
air shower. Using this method a complementary measurement to the settled technique of particle detectors 
is achieved. Several radio antenna arrays have been built or are going to be constructed, where LOPES in 
Karlsruhe served as the ground breaker for this cosmic ray detection technique. 
The Auger Engineering Radio Array AERA at the Pierre Auger Observatory in Argentina, presently being constructed, 
is the first array of a large scale application of the technique~\citep{ad-icrc09}. 
\begin{figure}[ht]
\begin{center}
\includegraphics[width= .45\textwidth ,angle=0]{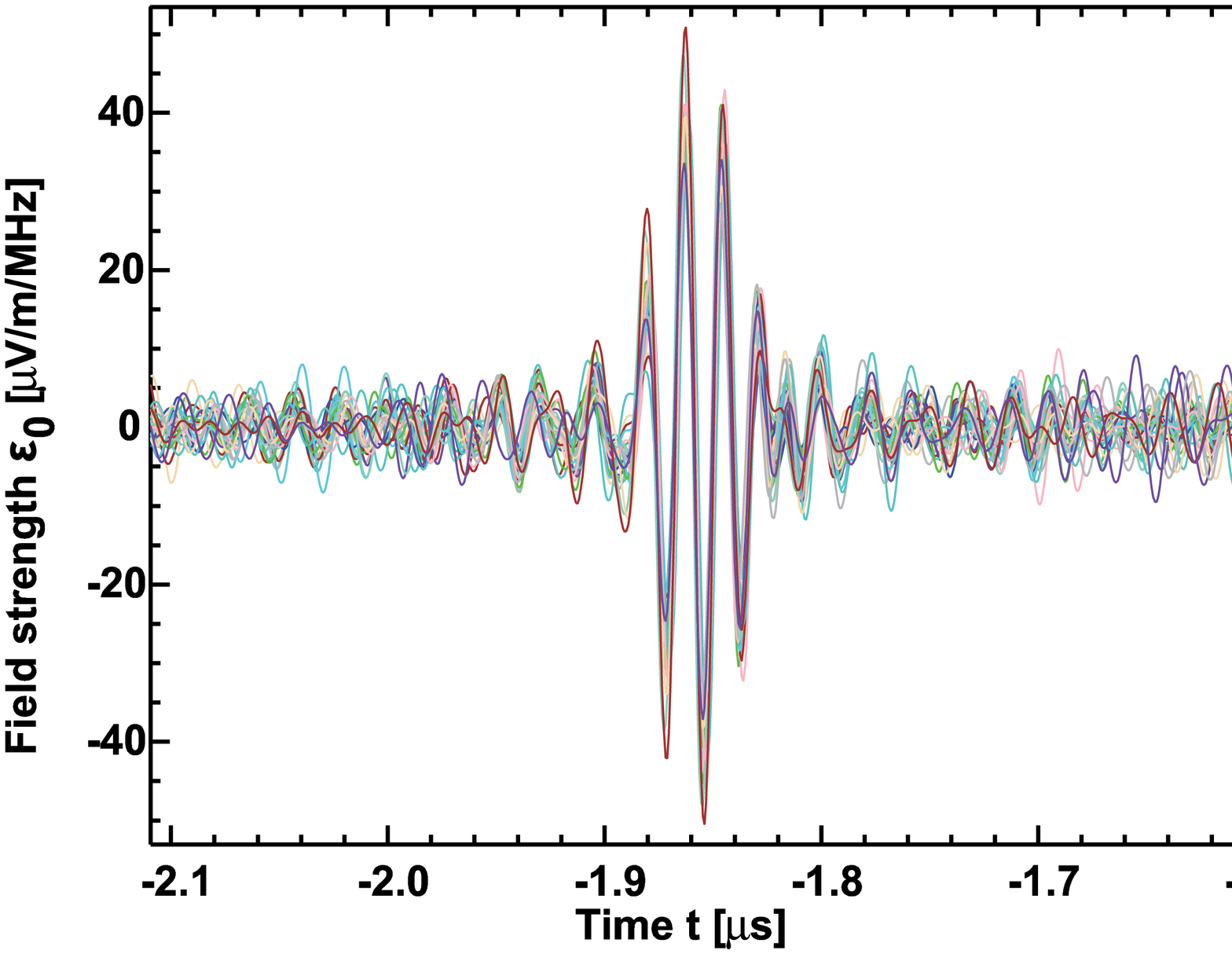} \quad
\includegraphics[width= .45\textwidth ,angle=0]{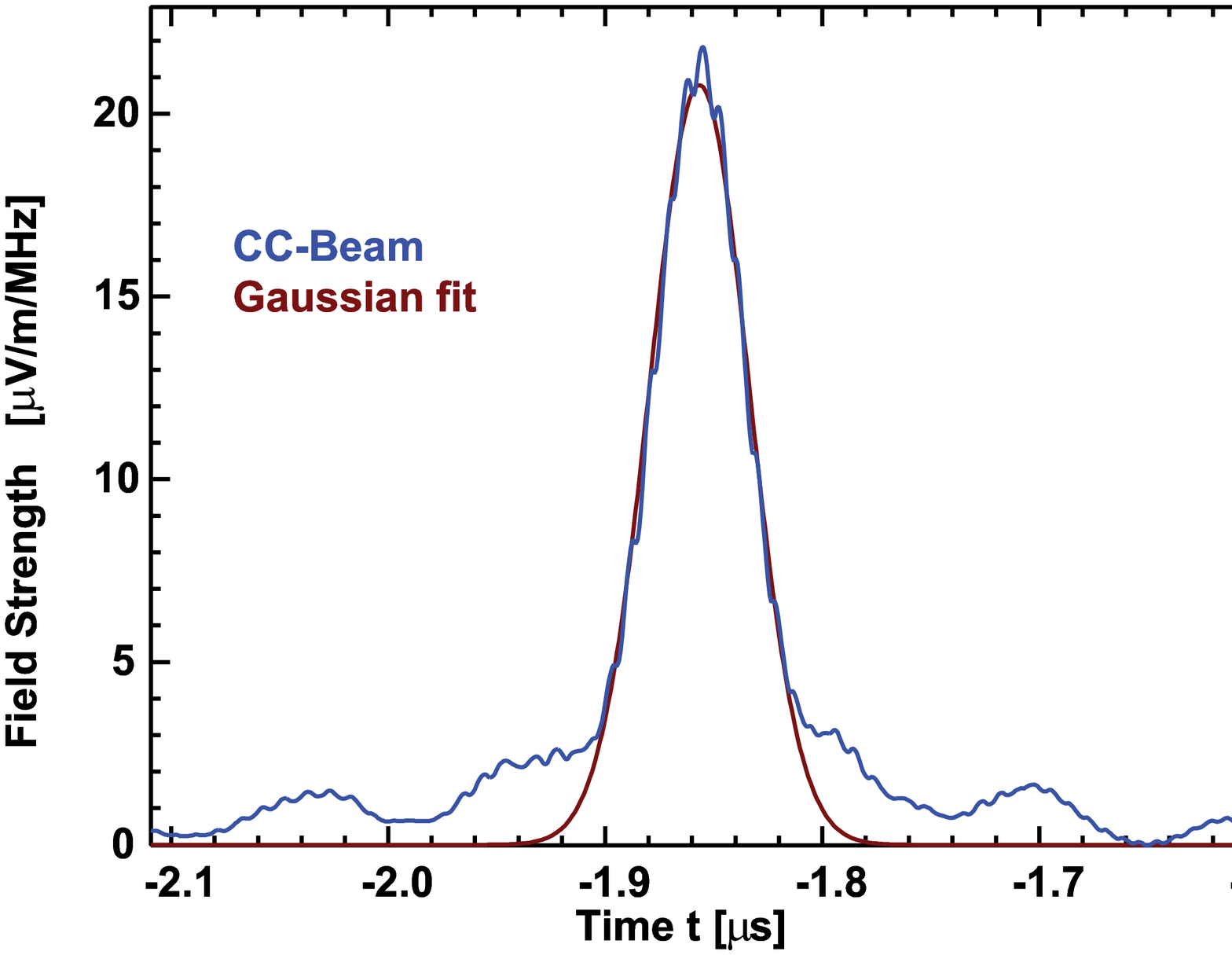}
\end{center}
\caption{Example of an EAS detection by the LOPES antenna array. 
Top: Bandwidth normalized field strength in $\mu$V/m/MHz vs. time in $\mu$\,s. The traces of all 
30 LOPES channels are shown. Bottom: Cross-correlation beam and Gaussian fit. 
The cross-correlation beam assigns time dependence of the averaged coherent part of the radio power 
received from the extensive air shower.}
\label{cc}
\end{figure}

By an interferometric standard analysis of a radio pulse measured by a dedicated radio antenna array 
several shower parameters can be determined, e.g. the arrival direction and the energy of the primary particle, 
or the shower maximum and therefore the mass of the incoming cosmic ray.  
To reconstruct these quantities several steps have to be performed in such an analysis, where
in the case of LOPES this is done as follows: 
To link the measured raw ADC counts with the field strength of the radio pulse 
the attenuation of the complete signal chain is investigated using an external source 
to calibrate the signal. 
The raw ADC counts for individual events are corrected for the electronic delay, attenuation, and dispersion 
by shifting the time traces and by multiplying the calibration factors. 
After that the narrow band noise is suppressed, where for this procedure the data are Fourier 
transformed from the time to the frequency domain. 
Although the pulses are filtered to a bandwidth where no broadcasting stations are operating in Karlsruhe, 
i.e. $40$ to $80$\,MHz, several narrow band noise sources can be seen in the spectra. 
Then, the reconstruction begins using the arrival direction of the cosmic ray measured by KASCADE-Grande 
as starting value. 
Taking into account the direction sensitive antenna characteristics, a 
so called beam forming is done. Such a beam forming is an iterative procedure to find the coherent part 
of the received signal in many antennas by shifting artificially the times until the direction 
is found where the coherent signal has its maximum. The beam forming in case of LOPES is based on the 
calculation of the cross-correlation value, which has a higher sensitivity to the coherent part of the signal
than, e.g., a simple power beam. 
Calculating the cross-correlation beam leads to an increase of the signal-to-noise ratio, because the 
coherent part of a signal is amplified and the incoherent part, the noise, is suppressed. 
This way the entire antenna array is used and even EAS signals that are not seen in individual antennas are 
reconstructed.
Since LOPES is sampling in the second Nyquist domain no information is lost, though the signal is recorded 
with a lower sampling rate than the highest frequency of the bandwidth to keep the recorded data small and 
save disk space. By an up-sampling procedure the original information is recovered.

Figure~\ref{cc} shows the time traces of all 30 LOPES antennas and the corresponding cross-correlation beam for 
an individual radio detected cosmic ray event. 
The time distribution of the cross-correlation beam values (top panel of Figure~\ref{cc})  
is fitted with a Gaussian function in 
order to obtain the radio EAS observables, where the most important is the field strength value at the maximum 
of the Gaussian function (bottom panel of Figure~\ref{cc}). 
This field strength value is directly correlated with the energy of the primary 
cosmic ray, only corrected for the geometry of the shower, i.e. arrival direction and position of 
incidence on observation level with respect to the antenna positions.

So far, LOPES has detected more than a thousand high-energy cosmic ray events serving as input on detailed analysis 
of the radio signal and the understanding of the radio emission in air showers in order to pave the way for 
large scale applications of the technique~\citep{Hau09}.  

\section{Atmospheric Electric Field during \hyphenation{Thunder-storms} Thunderstorms}
The exact measurement of the atmospheric electric field with a high time resolution is very important to 
provide information for the radio detection of cosmic ray induced air showers. The radiation in the radio 
regime is emitted by the propagating electrons and positrons of an air shower mainly due to their time 
dependent spatial (charge) separation in the Earth's magnetic field. 
The atmospheric electric field has an influence on the propagation of the charged particles of the shower 
and therefore on the radio emission. Only knowledge of the field  allows detailed studies of the 
influence of the atmospheric electric field on the radio emission. 
\begin{figure}[h]
\begin{center}
\includegraphics[width= .4\textwidth ,angle=0]{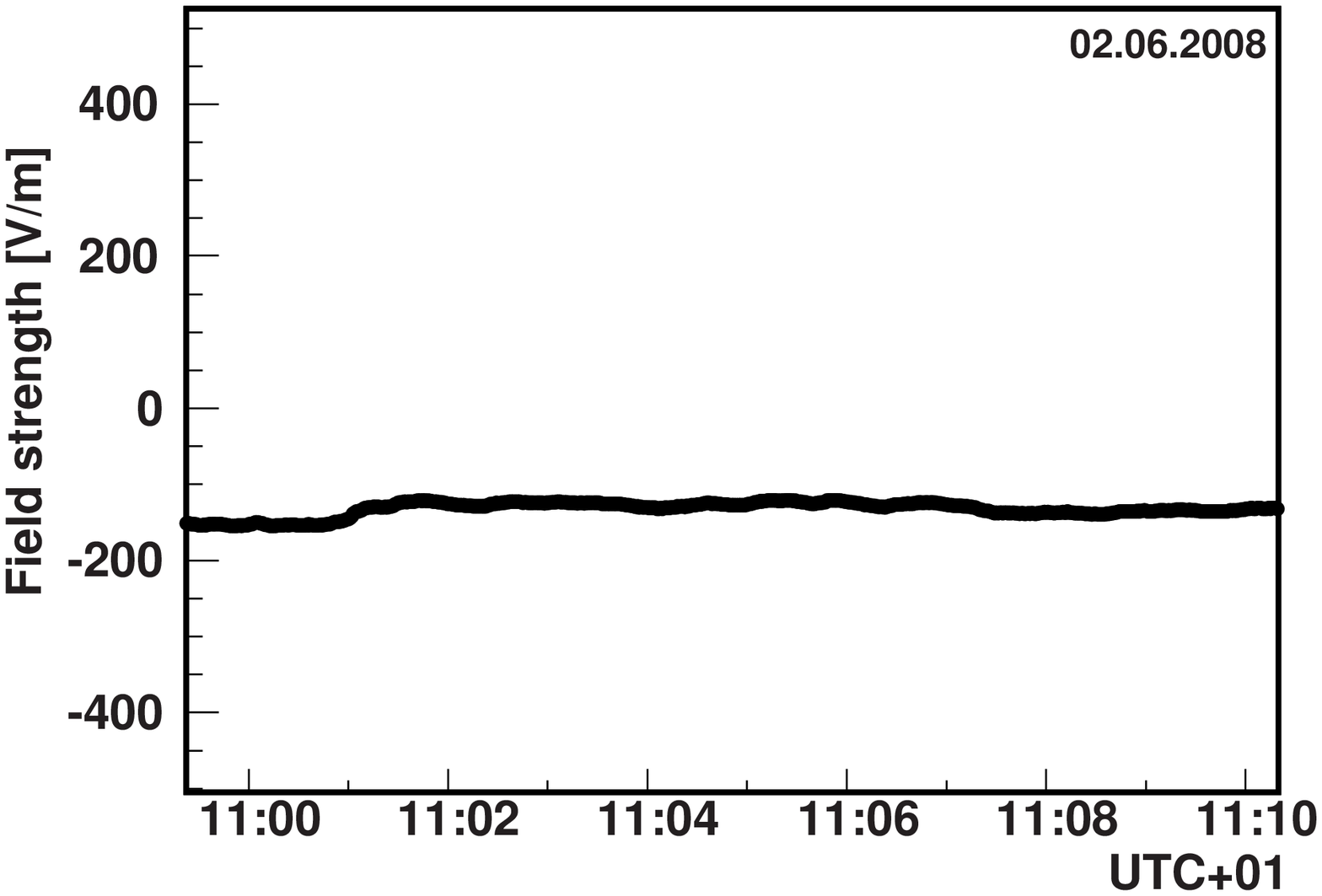}
\includegraphics[width= .4\textwidth ,angle=0]{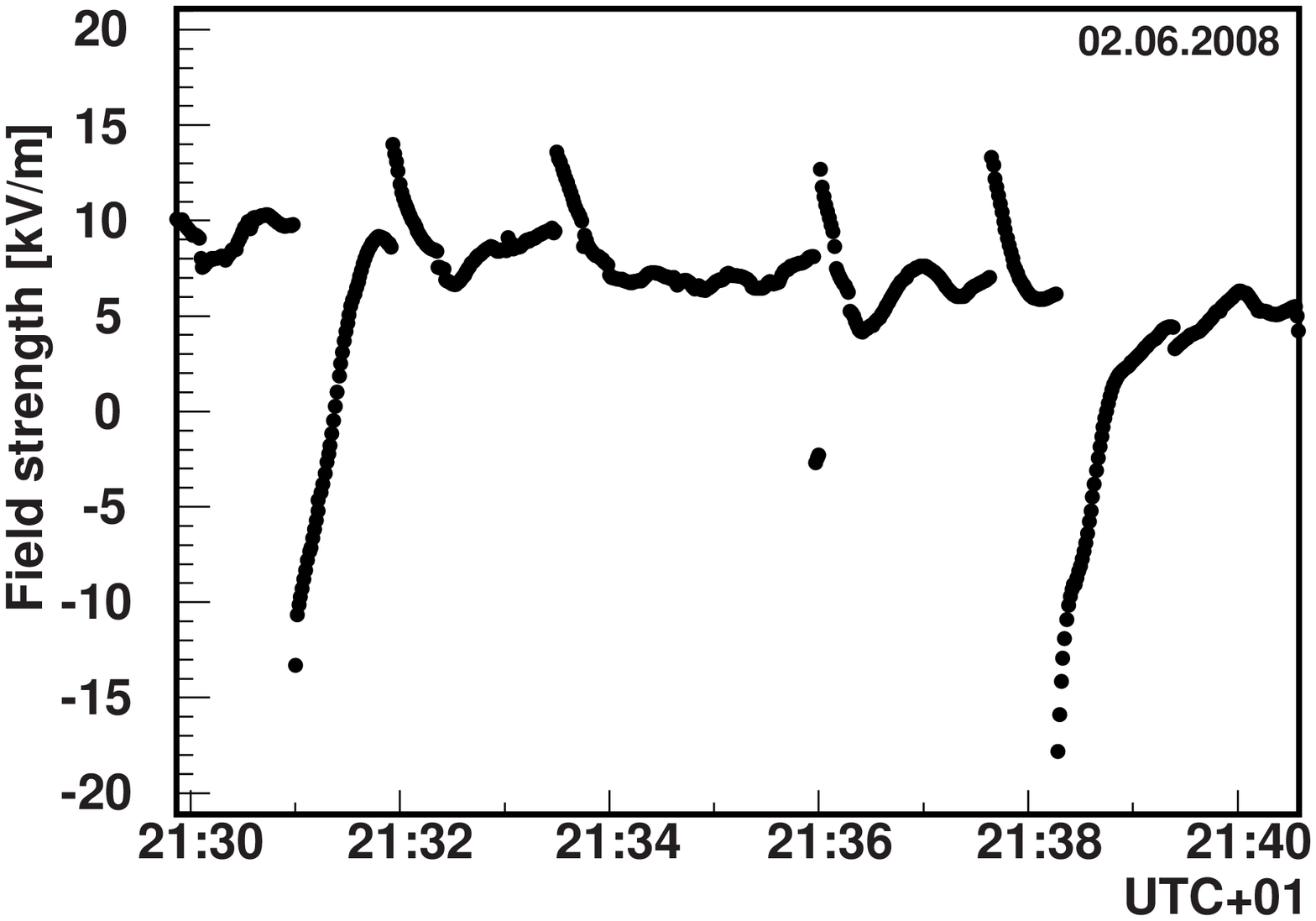} 
\end{center}
\caption{Electric field measured by an E-field mill. Top: During fair weather conditions. Bottom: During 
a thunderstorm, where discontinuities and jumps in the electric field strength are seen. In addition, 
a much higher field strength than during fair weather is measured.}
\label{efield1}
\end{figure}

To measure the atmospheric electric field a field mill is used. With such a device it is possible to 
record the vertical electrical field between the lowest cloud layer and the ground. The electric field 
strength gives clear evidences whether there are fair weather conditions or a close thunderstorm.

During fair weather conditions, the atmospheric electric field experiences only small changes between $-100$ 
and $-200$~Vm$^{-1}$. The amplitudes increase when rain clouds cross over but the changes are on large time 
scales. During thunderstorms the field strength can reach values up to $\pm 20$~kVm$^{-1}$ and there are sudden 
changes in the electric field occurring on a very short time scale, see figure~\ref{efield1}.
\begin{figure}[htb]
\begin{center}
\includegraphics[width= .25\textwidth ,angle=0]{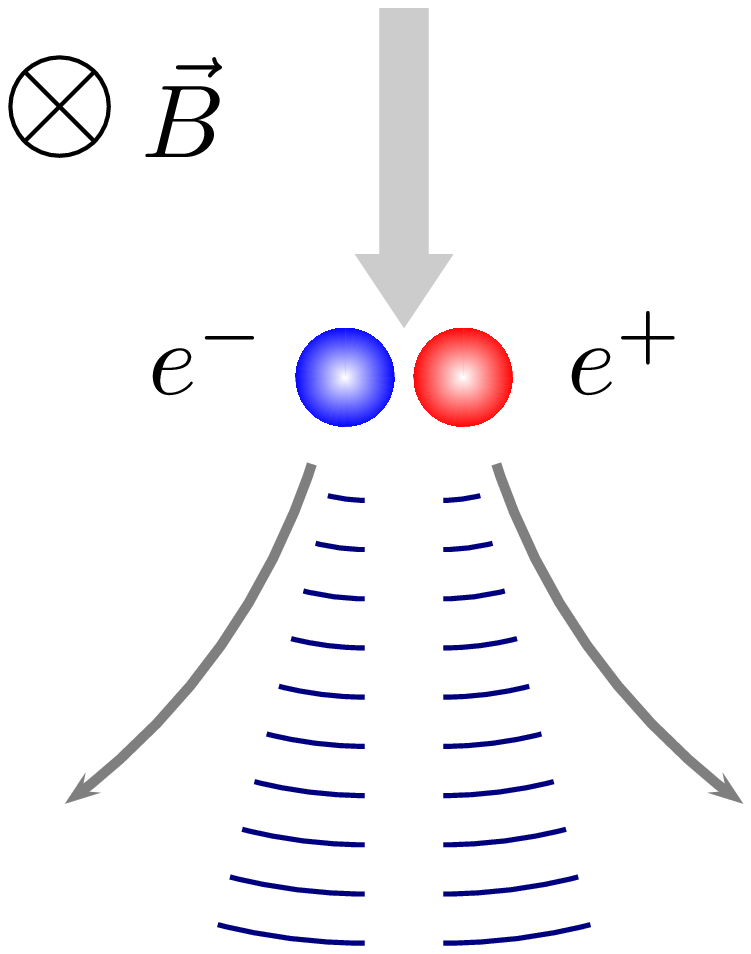} 
\includegraphics[width= .25\textwidth ,angle=0]{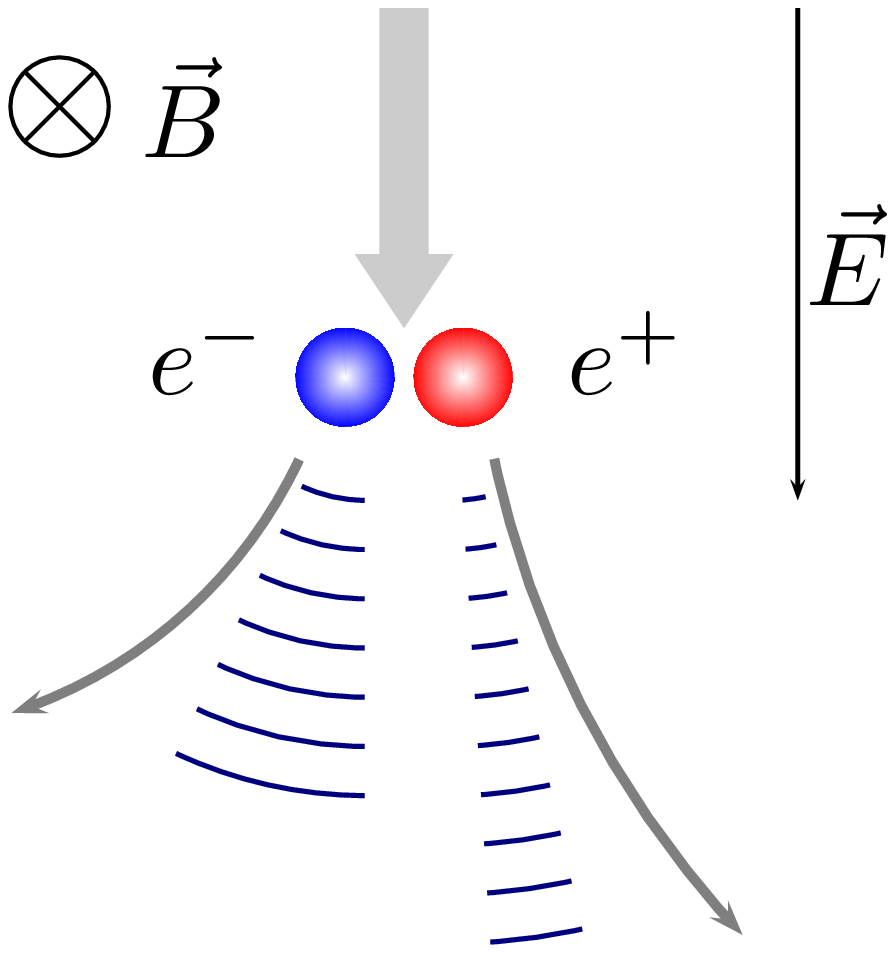}
\includegraphics[width= .25\textwidth ,angle=0]{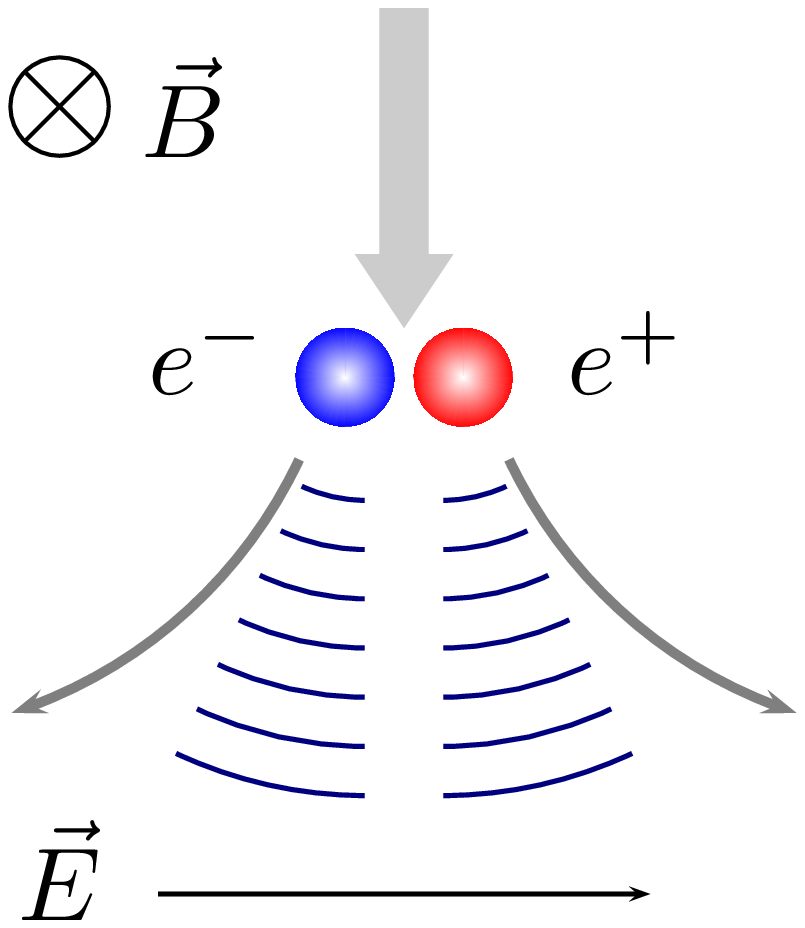}
\end{center}
\caption{Scheme of the influence of an additional electric field on the electrons and positrons of an air shower. 
On the left hand side no additional electric field is shown. In the middle, an electric field parallel to
the direction of the air shower is shown which leads to an acceleration of the positrons and a deceleration 
of the electrons. On the right hand side, an electric field which is perpendicular to the shower direction 
is shown which results in a stronger deflection of the electrons and positrons (from reference~\citet{buitink}).}
\label{efield2}
\end{figure}

\section{Influence of Atmospheric Electric Fields on Radio Detection of EAS}
\begin{figure}[htb]
\begin{center}
\includegraphics[width= .3\textwidth ,angle=-90]{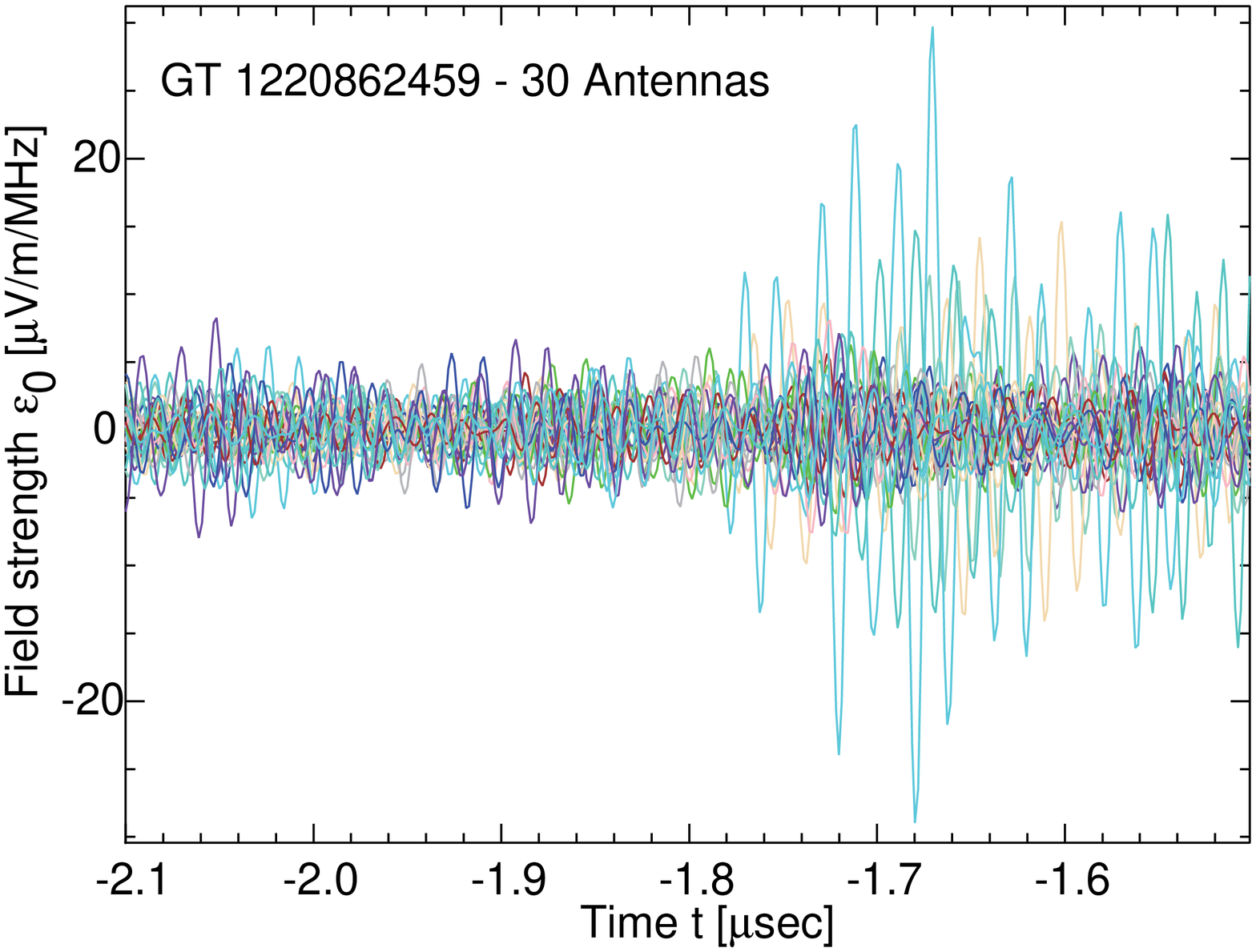} \\
\includegraphics[width= .3\textwidth ,angle=-90]{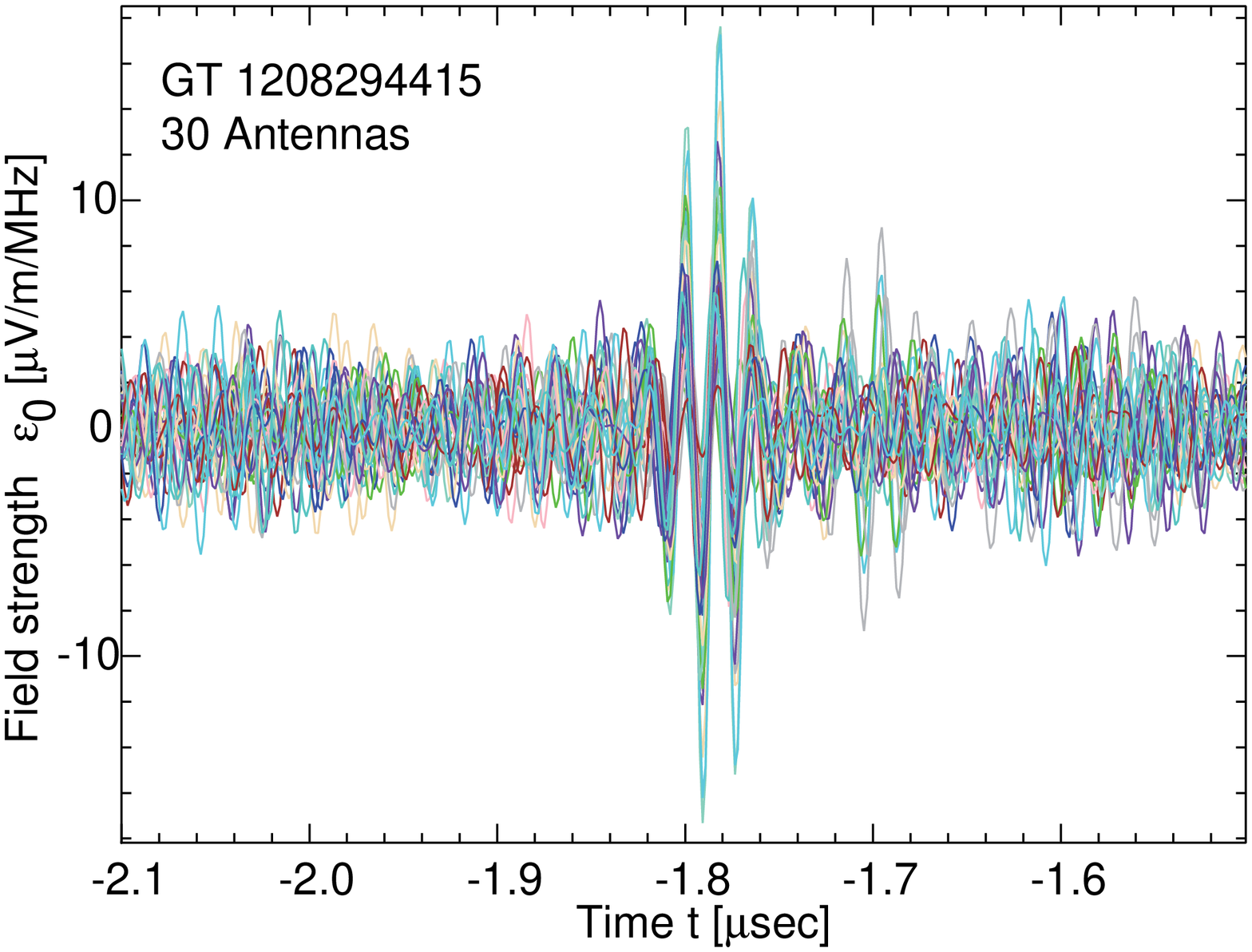}
\end{center}
\caption{Bandwidth normalized field strength in $\mu$V/m/MHz vs. time in $\mu$\,s for two events 
with similar geometry and primary energy. Top: Time traces of an 
event recorded during fair weather conditions, where no coherent signal can be seen. 
The increase in some of the traces starting at $-1.75\mu$\,s is assigned to detector noise from the KASCADE 
particle detectors. Bottom: Time traces of a similar event which was recorded during a thunderstorm. 
The amplified coherent signal from the direction of the air shower at $-1.8\mu$\,s is clearly visible.}
\label{twinevent}
\end{figure}

An additional or strongly varying atmospheric electric field can lead to different strengths and geometries 
of the radio emission from charged particles, in particular secondaries from a cosmic-ray air shower in the 
Earth's atmosphere~\citep{buitink,buitink2010}. 
In figure~\ref{efield2}, this effect is schematically shown. 
The electric field in thunderclouds, especially within the convective region, can reach values up 
to $\pm 100$\,kVm$^{-1}$. 
With radio antenna arrays, like e.g. LOPES, this effect can be seen in the observation 
of cosmic rays by recording an amplified or weakened radio signal during thunderstorms and extreme 
weather conditions. In order to detect a thunderstorm a field mill has been installed at the LOPES 
site~\citep{nehls} serving as a monitoring and veto device for the LOPES analysis. One example 
for such an amplification can be seen in figure~\ref{twinevent} where two events with very similar 
shower geometry and primary energy are shown. The first event was recorded during fair weather 
conditions with no observable radio signal in the trace which should occur  at about $-1.8\ \mu$s. 
This is expected for the low estimated primary energy by KASCADE-Grande observations of $5.4\cdot 10^{16}$~eV. 
The incoherent signal starting at $-1.75\ \mu$s is assigned to detector noise from the KASCADE particle detectors.
This air shower arrived at the KASCADE array with $\phi=110.39\,^{\circ}$ and $\theta=31.5\,^{\circ}$, 
where $\phi$ is the KASCADE reconstructed azimuth of the shower direction and $\theta$ the zenith angle.
The second event is an air shower with $\phi =110.35\,^{\circ}$ and $\theta =32.1\,^{\circ}$ 
and an even lower energy estimated to $4.3\cdot 10^{16}$~eV. The average distance of the antennas to the 
shower core is also in the same order for both events. 
In the time trace of this event a coherent radio pulse at -1.8~$\mu s$ is
observed, but for that energy no radio signal is expected. 
Most likely explanation for such a clear detection is the amplification of the EAS radio 
signal in the strong electric fields of the thundercloud present at that time. 

The energy threshold for triggering LOPES by KASCADE-Grande is lower than the possible detection 
threshold ($5\cdot10^{16}$\,eV) at the site of LOPES with its industrial environment and high noise level.  
Because of that only in a small fraction of the triggered events a radio pulse can be observed. 
In a fraction of $(0.96\pm 0.12)\cdot10^{-2}$ of the events recorded during fair weather conditions a 
coherent radio signal has been seen. 
In 2007 and 2008 approximately two full days of data were recorded at periods of thunderstorms. 
Because of these low statistics only few such corresponding fair weather - thunderstorm partner events 
could be analyzed. 
But the fraction of events that were recorded during thunderstorms and that show a coherent signal is 
with $(2.39\pm 0.27)\cdot 10^{-2}$ about a factor two to three higher than in fair weather conditions. 
This results in 81 events that are recorded during thunderstorms having a cross-correlation beam above 
threshold. It serves as a clear indication that strong atmospheric electric fields during thunderstorms 
have an influence on the radio emission of cosmic ray air showers and might more likely amplify the signal 
than attenuate it. For smaller atmospheric electrical fields there might be no big effect~\citep{buitink} 
but recent investigations have shown that strong atmospheric electric fields without thunderstorm that can 
occur in rain clouds also have an influence on the radio emission but the number 
of events recorded during such conditions are even lower~\citep{moses}. 

The aim of such studies is that by knowing the influence of electric fields on the radio emission in EAS 
the radio signal recorded during extreme weather conditions can be corrected for and therefore these periods 
are not lost for the analysis. 
This could lead to an uptime for the radio detection technique close to 100\%, which is important as 
the flux of high-energy cosmic rays is very low. 
For the time being the atmospheric monitoring is used as veto for the standard operation of radio antenna 
arrays such as LOPES.

\section{Radio Background during Thunderstorms}
\begin{figure}[htb]
\begin{center}
\includegraphics[width= .5\textwidth ,angle=0]{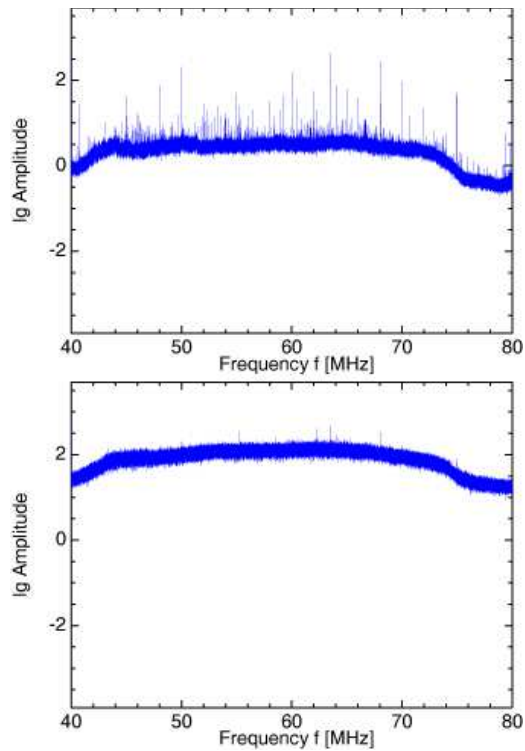} 
\end{center}
\caption{Logarithm of the mean amplitude of background frequency spectra in arbitrary units. Top: 
Background spectrum recorded during fair weather conditions, where a lot of narrow band noise sources are 
observed as small peaks. Bottom: Background spectrum recorded during a thunderstorm. The much higher broad band
background level exceeds nearly all narrow band noise peaks including the beacon signals needed for the 
event-to-event time calibration.}
\label{spec}
\end{figure}

To achieve the desired timing accuracy for beam forming reconstruction a reference antenna (beacon) was 
installed at LOPES. The beacon emits constant sine waves at 63.5, 68.1\,MHz and (since end 
of 2010) 53.1\,MHz which form a considerable background of the measurements during fair weather.  
During thunderstorms the general ambient background is much higher and in addition radio signal from 
lightning strikes contribute to the background. In figure \ref{spec} two average background frequency spectra 
are shown. The upper spectrum was recorded during fair weather conditions. The narrow band noise and the 
peaks of the beacon signals are clearly seen. The lower part of the figure shows a spectrum taken during 
thunderstorm conditions. Here, the beacon signals and other narrow band noise sources can hardly be 
identified over the high broadband radio emission.
As the beacon signals are overlapped in such periods the recorded events cannot be analyzed. 

Another aspect that makes it impossible to analyze individual events are the fact that the 
sensitive electronics needed to observe the weak radio signals from cosmic-ray air showers are 
saturated by the strong signals from lightnings. 

To be competitive with other EAS detection techniques it is aspired to trigger 
on the radio pulse only, in order to be independent from external triggers, e.g. by particle detectors. 
Within LOPES\textsuperscript{STAR}~\citep{asch} such a self-trigger technique is under development. 
The trigger condition is based beside other conditions on the pulse shape. 
To avoid wrong trigger decisions that arise from thunderstorms it is important to know the characteristics 
of non-EAS short pulses, e.g.generated by lightning.

\section{Problem of Lightning Initiation}
One of the most extreme weather conditions mankind can think of are thunderstorms and lightning. 
Although these appearances are known for a very long time and have been the focus of many studies, 
the mechanism that leads to the final electric breakdown is still not well known. 
The field strengths of the electric fields in thunderclouds are large, but too small for a classical 
breakdown. One mechanism that could explain how a breakdown can happen with smaller electric field 
strengths are relativistic runaway electron avalanches (RREA). 
These RREA occur at a certain critical field strength when the cross-section of the electron-electron 
interaction gets smaller which leads to ionization losses that are smaller compared to the energy gain 
at the critical field strength. This results in a net energy gain of the electrons and an increasing 
number of electrons. This is since the electrons coming from the ionization are also accelerated and 
again produce new unbound electrons. This mechanism can only take place when the number of unbound 
electrons that can be accelerated, the seed electrons, is high enough within the strong electric 
field of a thundercloud. A source that can provide these seed electrons are cosmic ray induced air showers. 
During an air shower development up to $10^{6}$ electrons and positrons can be generated on a very limited area, 
which is the location of the shower maximum, typically in a height of 3-8 km above seal level. 
These particles are then accelerated in the strong electric field of the thundercloud and lead to a RREA 
which results in a breakdown, called RRB, relativistic runaway breakdown. 
So, cosmic rays could be the initiator for lightning by providing the 
seed electrons for a RREA~\citep{dwyer2010}.

\section{Radio Signal from Lightning Strikes}
\begin{figure}[htb]
\begin{center}
\includegraphics[width= .6\textwidth , angle=0]{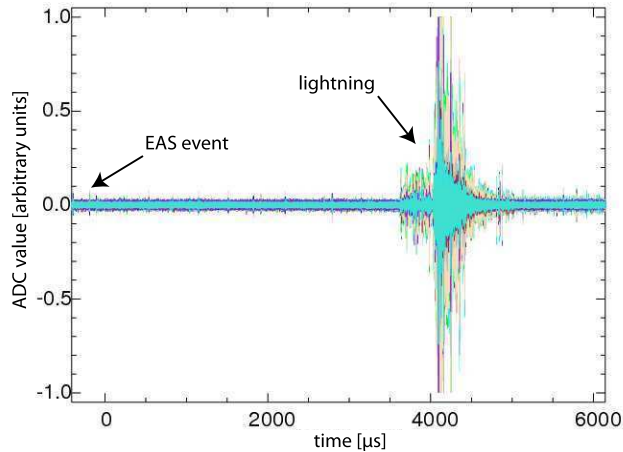} 
\end{center}
\caption{Example for time traces of the 30 LOPES antennas recorded in the thunderstorm mode for a 
KASCADE triggered air shower. 
Whereas the cosmic ray event triggering the readout is visible as small peak in the beginning of the 
traces, in this particular event a lightning strike occurs approximately 4 ms after the EAS. 
Structures in the radio emission of this lightning are very nicely resolved in all 30 antennas.}
\label{lightning}
\end{figure}

The jumps and discontinuities in the electric field are clear evidences for thunderstorms and 
are used at LOPES to change the data acquisition into the so-named thunderstorm mode~\citep{nehls}. 
During this special mode the recorded time traces are roughly eight times longer than the 
usual 0.82~ms and 6.55~ms of data taken for each triggered event, where the pre-trigger time 
of 0.41~ms remains the same. 
This is done to be able to look for timely extended lightning signals in the recorded traces
visible after the EAS signal, see an example in figure~\ref{lightning}. 
A discharging process like a lightning is always accompanied by strong electric fields in 
thunderclouds and always emits broadband electromagnetic radiation~\citep{lightning}. 
The radiation can be observed in the radio regime over long distances with antenna arrays  
originally designed to detect the radio emission from cosmic-ray air showers. 
The time structure of the signal can be very different depending on the distance and the nature of the 
discharging process. Figure~\ref{lightning} shows a lightning signal recorded by all antennas of the LOPES array.

Discharges located in clouds with altitudes between 5 and 20\,km above sea level can produce very short and 
strong pulses that are described as narrow bipolar pulses by~\citep{gurevich}. 
It could be an interesting question to investigate if such short pulses seen in figure~\ref{lightning} 
just in front of the large emission from the lightning strike are generated by relativistic runaway breakdowns (RBB)
and if such RRB's are always present at the initiation of lightning strikes. 
\begin{figure}[htb]
\begin{center}
\includegraphics[width= .75\textwidth ,angle=0]{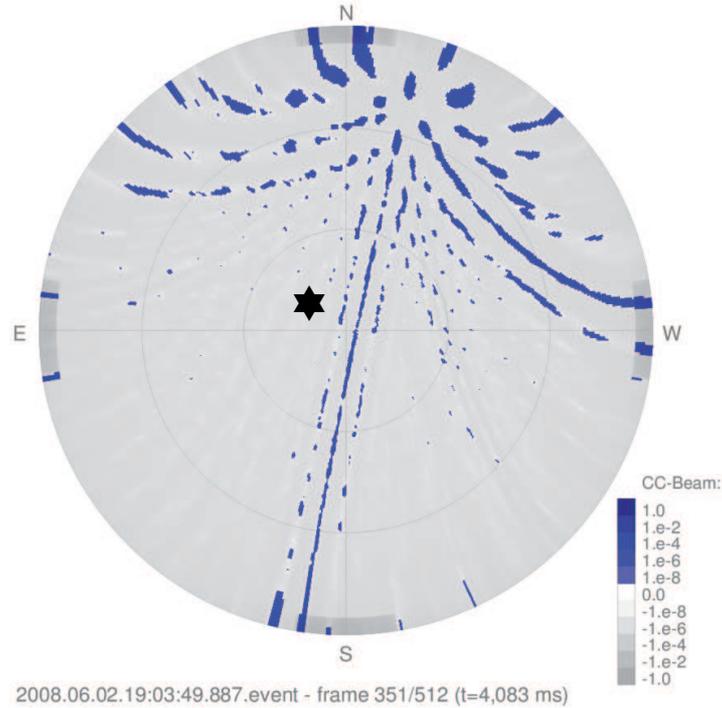} 
\end{center}
\caption{This figure shows a skymap of a lightning recorded with LOPES, where for each cell 
the amplitude of the cross-correlation beam is displayed.  The whole sky is shown with 
the zenith in the center of the plot and the horizon at the edge. The lightning can bee seen as the 
strong signal in the North-Northwest (circle). The signals spread over the whole sky are due to the grating lobes 
of the antenna array which is a well understood artifact of skymapping in astronomy and only dependent 
on the geometry of the antenna positions on ground. 
The event corresponds to the traces shown in figure~\ref{lightning} and the star assigns the location 
of the detected EAS in this event.}
\label{skymap}
\end{figure}

Discharges within or between clouds or from cloud to ground result in signals that are lasting 
longer than RRBs or EAS signals (see figure~\ref{lightning}). 
This gives us the possibility to calculate for many successive time slots and in every direction of the sky 
the cc-beam value and to combine them into a skymap. Typically the cross-correlation beam is used, since the 
signal from lightning strikes is coherent, but if desired also the power beam 
(averaged power of all antennas in a certain direction) is used to generate the skymap. 
By this procedure the lightning development is observed and the location of origin, 
the type and the direction of the lightning is reconstructed. 
For the principles of lightning observation using radio interferometers see also~\citep{rhodes}. 
For example the lightning strike displayed in figure~\ref{skymap} (which corresponds to the time traces 
shown in Figure~\ref{lightning}) is a cloud-to-cloud lightning as the track does not reach the horizon.
This example shows the capabilities of EAS radio antenna arrays to investigate in detail the
radio emission from individual lightning discharges.

\section{Correlations of EAS and Lightnings}
To study the possibility of cosmic rays causing lightning strikes there are two correlations to investigate,
the time and the spatial connections of air showers and lightning.

\begin{figure}[t]
\begin{center}
\includegraphics[width= .7\textwidth]{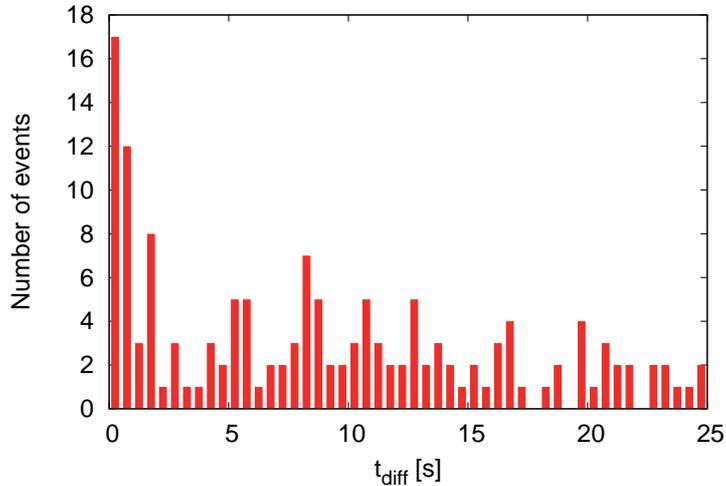} 
\end{center}
\caption{Time correlation between lightning and air-shower events. Shown is the time difference 
between cosmic-ray event and lightning, where the lightning is detected by an electric field mill.}
\label{timecorr}
\end{figure}

To investigate the timely correlation, at LOPES a lightning is detected by the E-field mill, where a jump 
in the electric field corresponds to a discharge process.
If cosmic-ray air showers induce these lightning strikes they should be observed by KASCADE-Grande or 
LOPES shortly before the lightning. 
In figure~\ref{timecorr}, the time differences between jumps in the electric field with detected 
cosmic ray events is shown. 
The time binning of one second is determined by the time resolution of the electric field mill used 
at the LOPES site. 
In order to investigate systematically introduced uncertainties, similar plots with same dataset 
but artificially introduced time delays to scramble the possible correlation were produced.
These studies resulted in only statistically insignificant enhancements at $t_{diff} = 0$. 

The time resolution of the electric field mill is too low for such an analysis, but
is high enough to determine whether there is a thunderstorm or not. To see a time correlation 
between cosmic-ray air showers and lightnings, a better time resolution is necessary. 
This can be provided by LOPES, but the statistics for such an analysis is still too low at LOPES. 
In addition, the area where EAS can be detected is much smaller than
the sensitive area for lightning strikes, which also worsen the search for time correlations.

A more promising  correlation study between cosmic rays and lightning is the search for a combined 
spatial and time correlation. 
To study this it is essential to detect lightnings with a high spatial resolution and to detect 
cosmic-ray air showers with a high efficiency and also a good spatial resolution. 
For the lightning detection a LOPES like array can be used as a combined lightning mapping and EAS detection 
array. 
The direction of the lightning can be determined by calculating the cross-correlation beam in every direction 
of the sky. A calculated skymap is shown in figure \ref{skymap}, where an intracloud discharge 
can be seen. 
A star marks the direction of the air shower arriving shortly before the lightning. 
In this event and also in others no correlation could be observed.  
To improve the searching for spatial correlations a better detection of the lightning and especially the 
lightning development is needed. Not only the direction where the lightning happened is of interest but 
also the path of the lightning and whether a cosmic-ray air shower passed there at the start of the 
lightning or somewhere near. 
The path of the air shower can be observed very well and reconstructed with KASCADE-Grande and LOPES. 
The development of the lightning is difficult to reconstruct with the given instruments since LOPES was 
not designed for that kind of studies and still the covered area is too small. 

\section{Future Investigations}
The LOPES experiment, which will stop operation by 2011, experienced some changes in hardware, but still 
will be able to measure in the thunderstorm mode in the summer period 2011. Now, LOPES consists of 10 
antenna stations with each having three simple dipole antennas sensitive to the  east-west, north-south, 
and vertical polarization direction, respectively.  
This makes the antenna array more sensitive, in particular to signals arriving from more inclined sources, 
i.e. from the horizon. 
In addition, electric field mills, an additional small lightning mapping array, and three antennas sensitive 
to radiation in the kHz frequency range will be active during the thunderstorm periods. 
In particular the kHz-information can be very useful for a combined 
multi-parameter observation of air-shower and lightning, as lightning strikes are strong emitters in 
these frequency domains. The KASCADE-Grande detector stations will be used to study 
the particle densities during a thunderstorm. 

Among different possible techniques, radio observations with dedicated antenna arrays 
are most suitable to follow the lightning development in time and space with 
highest resolution. 
Such lightning mapping arrays (LMA)~\citep{krehbiel} are devices especially designed to better measure 
the initiation and generation of lightning and can be used - when co-located to air-shower experiments - 
to investigate the influence of thunderstorms and lightning on the air shower detection by particle detectors.
The "Lightning Air Shower Study" project LASS as a part of the Pierre Auger Observatory~\citep{abraham} 
is in this sense the next generation experiment for lightning and thunderstorm 
investigations in combination with an air-shower experiment.
LASS will help to better understand the development of lightning as well as to investigate 
the influence of thunderstorms and lightning to the air shower detection by particle detectors, 
by fluorescence telescopes, and by radio antenna arrays such as LOPES or AERA, where AERA will be 
combined with the lightning mapping array.  

By deploying different types of radio antennas and highly sensitive electric field mills a lightning can be 
observed with a spatial resolution of around $10\,$m and time resolution of about $40\,$ns~\citep{thomas}.
Data from the LMA stations can then be combined to provide three-dimensional images of the lightning channels, 
including lightning initiation positions and times of occurrence. 
Each LMA station records peak signal magnitude and time in every $80\,\mu$s interval in a 
quiet 6 MHz VHF band at typically $63$ MHz. 

The goals of such a project are the determination of a possible correlation between lightning 
and cosmic-ray air showers, to study the effects of strong electric fields on the radio emission of EAS, 
to study the influence of strong electric fields or thunderstorms on the particle component of EAS and 
to investigate whether distant lightnings harm the performance of the fluorescence detectors. 
For that purposes the installation within an existing EAS experiment, such as the Pierre Auger Observatory, 
is ideal. Both projects could profit from each other as lightning can also affect the air-shower measurements 
by producing a broadband light flash that can harm fluorescence measurements over long distances. 
The X- and gamma rays produced by lightning strikes can irradiate the particle detectors and cause background. 
The moving charge in a lightning causes also emission in the MHz range which causes additional radio background. 
The strong atmospheric electric fields during thunderstorms can seriously have an influence on the 
particle distributions of charged particles from an EAS in the atmosphere, which strongly affects the correct 
reconstruction of this air shower.

\section{Conclusion}
The LOPES experiment has provided the proof of principle for the radio detection technique of high-energy 
cosmic rays and has made key contributions to the understanding of the radio emission physics such as the
predominantly geomagnetic origin of the emission. LOPES significantly contributes to the calibration 
of the shower radio emission in the primary energy range of $10^{16}\,$eV to $10^{18}\,$eV. 
I.e., is investigating in detail the correlation of the measured field strength with the primary  
cosmic ray parameters, in particular the arrival direction, the energy and the mass.
The radio detection technique  has a high reliability in all but the most extreme weather conditions, as 
for example the reconstructed energy of an air shower is influenced by electric fields during thunderstorms, 
but not at normal weather conditions leading to a larger fraction of events with a detected coherent signal.
Therefore it is mandatory for antenna air-shower arrays to monitor the electric field of the atmosphere.

Lightning strikes are one of the most extreme weather appearances that can occur in nature. 
Although they have been analyzed for a very long time, the basic mechanism that leads to the final 
breakdown is still not known. 
By observing lightning strikes in the radio regime the strikes can be observed looking through the 
thunderclouds since these are transparent for radio waves. This gives huge advantage to the studies of lightning 
in the radio regime over optical observations. Clouds are opaque for optical transmission which makes it 
impossible to observe lightning strikes properly. 
With the radio antenna array LOPES lightning strikes could 
be observed, although LOPES was not especially designed for that. 
LOPES was built within the existing cosmic-ray air shower experiment KASCADE-Grande. 
This allowed first low-level studies of correlations between cosmic rays and lightning strikes, by analyzing 
the air-shower properties given by KASCADE-Grande and the geometry of the lightning strike observed with LOPES.
These first analyses showed no significant correlation between cosmic rays and lightning strikes. 

The LOPES results are interesting and verified the capabilities of such arrays in lightning studies, but because 
of the drawbacks of a small area covered and short time traces recorded a final conclusion is not possible yet.  
To draw a final conclusion whether cosmic rays induce lightning strikes or not, more sophisticated studies with devices 
that are specially designed for that purpose need to be performed. This will be realized by future 
experimental devices, such as LASS, where high tech air shower analyses and lightning detection over a huge area 
can be combined. 

\section*{Acknowledgment}
LOPES and KASCADE-Grande have been supported by the German Federal Ministry of Education and Research.
KASCADE-Grande is partly supported by the MIUR and INAF of Italy, the Polish Ministry of Science and 
Higher Education and by the Romanian Authority for Scientific Research CNCSIS-UEFISCSU 
(grant IDEI 1442/2008). Part of this research has been supported by grant number VH-NG-413 of the 
Helmholtz Association. 
D.H. and A.H would like to thank Paul Krehbiel and his colleagues from New Mexico Tech, Joe Dwyer from 
Florida State University, and in particular Bill Brown, Colorado State University, for the initiation of 
the LASS project and the possibility to participate on these coming efforts. 
D.H. would like to thank the organizers of the TEPA conference held in Armenia, in particular Ashot Chilingarian, 
for the possibility to discuss the topic in an excellent atmosphere.


\begin{thebibliography}{}


  \bibitem[Abraham et al.(2004)]{abraham} J.~Abraham et al. - Pierre Auger Collaboration,
   NIM A, 523, 50, 2004.
  \bibitem[Antoni et al.(2003)]{kascade} T.~Antoni et al. - KASCADE Collaboration,
   NIM A 513, 429, 2003. 
  \bibitem[Apel et al.(2010)]{grande} W.~D.~Apel et al. - KASCADE-Grande Collaboration,
   NIM A, 620, 202-216, 2010.
  \bibitem[Asch(2009)]{asch} T.~Asch, FZKA report  7459,
   Forschungszentrum Karlsruhe 2009.
  \bibitem[Buitink et al.(2010)]{buitink2010} S.~Buitink, T.~Huege, H.~Falcke, J.~Kuijpers,
   Astroparticle Physics, 33, 296-306, 2010.
  \bibitem[Buitink et al.(2007)]{buitink} S.~Buitink et al. - LOPES Collaboration,
   A\&A, 467, 385, 2007.
  \bibitem[Dwyer et al.(2009)]{dwyer} J.~R.~Dwyer, M.~A.~Uman and H.~K.~Rassoul,
   J. Geophys. Res., 114, D09208, 2009.
  \bibitem[Dwyer(2010)]{dwyer2010} J.~R.~Dwyer,
   J. Geophys. Res., 115, A00E14, 2010.
  \bibitem[Ender et al.(2009a)]{mosesICRC} M.~Ender et al. - LOPES Collaboration,
   Proceedings of the 31st ICRC, Lodz, Poland, 2009.
  \bibitem[Falcke et al.(2005)]{Falcke05} H. Falcke et al. - LOPES Collaboration,
   Nature 435, 313, 2005.     
  \bibitem[Gurevich \& Zybin(2004)]{gurevich} A.~V.~Gurevich, K.~P.~Zybin,
   Physics Letters A 329, 341-347, 2004.
  \bibitem[Haungs et al.(2003)]{haungs2003} A.~Haungs, H.~Rebel, M.~Roth,
   Reports on Progress in Physics 66, 1145, 2003.
  \bibitem[Haungs et al.(2009)]{Hau09} A.~Haungs et al. - LOPES Collaboration,
   NIM A 604, 1,  2009.
  \bibitem[Nehls(2008)]{nehls} S.~Nehls,
   FZKA report 7440, Forschungszentrum Karlsruhe 2008.
  \bibitem[Rakov \& Uman(2005)]{lightning} V.~Rakov, M.~Uman,
   ISBN 0-521-03541-4, CAMBRIDGE Univ. Press, 2005.   
  \bibitem[Rhodes et al.(1994)]{rhodes} C.~T.~Rhodes, X.~M.~Shao, P.~R.~Krehbiel, R.~J.~Thomas, and C.~O.~Hayenga,
   J. Geophys. Res., 99, 13,059-13082, 1994.
  \bibitem[Rison et al.(1999)]{krehbiel}W.~Rison, R.~J.~Thomas, P.~R.~Krehbiel, T.~Hamlin, and J.~Harlin, 
   Geophys. Res. Lett., 26(23), 3573-3576, 1999.
  \bibitem[Schroeder et al.(2010)]{SchroederArena} F.~G.~Schroeder et al. - LOPES Collaboration,
   submitted to Nucl. Instr. Meth. A - ARENA proceedings (2010b),arXiv:1009.3444.
  \bibitem[Thomas et al.(2004)]{thomas} R.J.Thomas,  P.~R.~Krehbiel, W.~Rison, S.~J.~Hunyady, W.~P.~Winn, T.~Hamlin and J.~Harlin,
   J. Geophys. Rsch., 109, D14207, doi:10.1029/2004/JD004549, 2004.  
  \bibitem[van den Berg et al.(2009)]{ad-icrc09} A.~van~den~Berg  et al. - Pierre Auger Collaboration,
   Proceedings of the 31st ICRC, Lodz, Poland, 2009.


\end{thebibliography}
\end{document}